\documentstyle[preprint,aps,floats]{revtex}
\input epsf.tex
\def\DESepsf(#1 width #2){\epsfxsize=#2 \epsfbox{#1}}

\def\appendix{\par
 \setcounter{section}{0}
 \setcounter{subsection}{0}
 \def\thesection{Appendix \Alph{section}}
 \def\thesubsection{\Alph{section}.\arabic{subsection}}
 \def\theequation{\Alph{section}.\arabic{equation}}
 \setcounter{equation}{0}}
\newcommand{\be}{\begin{equation}}
\newcommand{\ee}{\end{equation}}
\newcommand{\bear}{\begin{eqnarray}}
\newcommand{\eear}{\end{eqnarray}}

\begin{document}
\preprint{ \vbox{ \hbox{ {\bf hep-ph/0211401}} }} \vskip 0.5cm
\title{A Flexible Parametrization of CKM matrix via \\
Singular-Value-Decomposition Method}

\author{
C.~S.~Kim\footnote{cskim@mail.yonsei.ac.kr},~~ and~~
J.~D.~Kim\footnote{jdkim@phya.yonsei.ac.kr}}

\address{
Department of Physics and IPAP,
Yonsei University, Seoul 120-749, Korea} \vskip -0.75cm

\maketitle
\begin{center}\today\end{center}
\begin{abstract}
\vskip-5ex 
\noindent We investigate a flexible method in which we can test the unitarity of
the quark flavor mixing matrix step-by-step. Singular-Value-Decomposition (SVD)
techniques are used in analyzing the mixing matrix over a broader
parameter region than the unitarity region. Unitary constraints
make us extract CP violating properties without any specific
parametrization when the magnitudes of at least three mixing matrix
elements in three generation quark mixing are given. This method
can also be  applied to the analysis of lepton
flavor mixing, in which only a few moduli are presently measured.
\end{abstract}

\newpage
\parskip 12pt

\section{Introduction}

The Cabibbo-Kobayashi-Maskawa (CKM)~\cite{Cabibbo,KM} matrix  makes us possible to
explain all flavor changing weak decay processes and CP violating
phenomena up to now. Unitarity of the CKM matrix in the standard model (SM)
is a unique property that we cannot loosen. We can use any parametrization of
the CKM matrix as long as its unitarity is conserved. The original
parametrization for three generation  quark mixing is the
Kobayash-Maskawa (KM) parametrization. The standard
parametrization proposed by Chau and Keung~\cite{Chau,PDB} is the
product of three complex rotation matrices which are characterized
by the three Euler angles, $\theta_{12},~\theta_{13},~\theta_{23}$
and an CP--violating phase $\delta_{13}$. More widely used one is the
Wolfenstein parametrization~\cite{Wolfenstein}, which was suggested
as a simple expansion of the CKM matrix in terms of the four
parameters: $\lambda,~A,~\rho$ and $\eta$. It has been also known that
the CKM matrix for the three-generation case can be parameterized
in  terms of the moduli of four of its
elements~\cite{Bjorken}. This
four-value-KM (4VKM) parametrization is rephasing invariant and
directly related to the measured quantities. In three generation
case we always need four independent parameters to define a unitary
$3 \times 3$ matrix, as explained, eg.
$\theta_{12},~\theta_{13},~\theta_{23}$ and $\delta_{13}$, or
$\lambda,~A,~\rho$ and $\eta$ or even only moduli of any four
independent elements of the matrix.

The 4VKM parametrization has several advantages over the other
parametrization. This parametrization doesn't need any specific
representations for the mixing angles as long as the CKM is unitary,
and no ambiguity over the
definition of its complex CP phase is present above all.
Secondly, the Jarlskog invariant quantity
$J_{cp}$ and non-trivial higher invariants can be reformulated as
functions of moduli and quadri-products~\cite{Branco2}.
However, in the 4VKM parametrization
initial four-moduli input values should be fixed by experiments.
Once we set four moduli to specific values, remaining five moduli of
mixing elements are automatically fixed and we may lose some characteristic
effects from interplaying between the moduli. In a conceptual point of
view it is better if we can reduce number of {\it a priori}
experimental input values. This paper presents a novel
parametrization in which we start with three-moduli input values.
Through simple algebraic relations we
can determine remaining six moduli of mixing elements.
With more broader parameter space we can
check the compatibility between measured values of mixing elements
and their unitarity properties step-by-step.

Many groups have made global fits and numerical works on CKM
matrix elements with conventional representations which satisfy
unitarity \cite{CSKIM}. One of the
problems in these conventional parameterizations
is that they are {\it fully} and {\it completely} unitary and are
not flexible to include possible non-unitary properties resulted from unknown new
physics. Therefore, it is a complicate task to make a step-by-step test
to check the unitarity with experimental data if you use a unitary
parametrization. In the following,
we present three extended definitions for the unitarities of mixing matrix
$V$ in the order of the strength of the constraints:
\begin{itemize}
\item{\underline{Weak Unitary Conditions (WUC)}:} We define that
the mixing matrix $V$ is weak unitary if it satisfies
\begin{equation}
\sum_{\alpha} |V_{i\alpha}|^2 =
\sum_{j} |V_{j\beta}|^2 = 1
\mbox{  for all } i=u,c,t, \mbox{ and } \;\;\; \beta=d,s,b.
\label{WUC1}
\end{equation}
These constraints appear to be well satisfied experimentally for
the three generation case, and we start from this. Actually it was
pointed out that there is an apparent functional violation in the
available data:
$|V_{ud}|^2+|V_{us}|^2+|V_{ub}|^2<1$~\cite{Hocker}.
For such a case with $\sum_{\alpha} |V_{u\alpha}|^2 =a <1$, we can
easily generalize our method, and we just start with this new condition.
\item{\underline{Almost Unitary Conditions (AUC)}:} In addition to
the constraint Eq.~(\ref{WUC1}), if the following constraints are
satisfied
\begin{equation}
\sum_{\alpha,i\neq j} V_{i\alpha}^* V_{j\alpha} =
\sum_{j, \alpha\neq\beta} V_{j\alpha}^* V_{j\beta} = 0
\mbox{  for some parts of } i,j=u,c,t, \mbox{ and } \;\;
\alpha,\beta=d,s,b,
\label{AUC1}
\end{equation}
let us call the mixing matrix almost unitary. Some combinations,
which do not satisfy Eq. (2), may not make closed triangles, and
may have different areas even though making closed triangles. We
have no specific models in which the mixing matrix satisfies this
almost unitary conditions. Therefore, we will not consider the
case with AUC.
\item{\underline{Full Unitary Conditions (FUC)}:} This corresponds
to usual unitarity in which Eqs.~(\ref{WUC1}),~(\ref{AUC1}) are
satisfied {\it for all the indices}. All six unitarity triangles
from Eq.~(\ref{AUC1}) have the same areas.
\end{itemize}

In Sec. II,  we propose an alternative and more flexible
parametrization of the CKM matrix in terms of the three moduli and
the one independent parameter, which is induced by
the singular-value-decomposition (SVD) method. We describe
how to get the new parametrization of the CKM matrix by using the
SVD method in the three-generation case. Unlike from the
previous parametrization with four moduli
\cite{Bjorken}, we have more
flexible leverage to test the unitarity step-by-step. We start
with only three moduli rather than four moduli, and the remaining
one can be adjusted depending on the condition of the unitarity,
which we apply, $i.e.$ WUC or FUC.
In Sec. III,
we analyze the CKM matrix numerically with our parametrization
with the SVD method. Conclusions are also in Sec. III. Appendices A--B
include details about the SVD method.


\section{New parametrization of the CKM matrix by the SVD method}

We start with a definition in such a way that it satisfies the
weak unitary conditions, Eq. (1): we have six constraint
equations for the 3 generation mixing.
These constraints are only parts of unitarity conditions and
the introduced mixing matrix $V$ may not be fully unitary. We
study this explicitly with three generation quark flavor
mixing matrix $V$ in their absolute values and choose three
independent moduli as starting input parameters. Explicit analysis
depends on the choice of three input parameters.
We consider the case with:
\begin{itemize}
\item{Our Choice (Set--A)}:
Input parameters  $|V_{us}|,~|V_{ub}|,~|V_{cb}|$.
\end{itemize}
We can also choose different sets of input parameters, as examples:
\begin{itemize}
\item{Set--B}: Input parameters  $|V_{ud}|,~|V_{us}|,~|V_{cd}|$,
\item{Set--C}: Input parameters  $|V_{us}|,~|V_{cs}|,~|V_{cb}|$.
\end{itemize}
Mathematically three parametrizations of Set--A,B,C are all equivalent if
three input values of each set are independent one another and
all equally precisely measured.
However, in reality, the upper-left $2\times 2$ part of CKM
matrix is {\it approximately} unitary and only one independent variable is
dominantly evident, for example, the parameter $\lambda$ in
the Wolfenstein parametrization or the Cabibbo angle
$\sin{\theta_c}$. Therefore, Set--B would be the worst choice
for numerical analyses.
For our choice of Set--A,
the three inputs are all off-diagonal and independent each other,
and all three values can be determined by three semileptonic decays,
in which new physics contributions are severely suppressed.
Therefore, we select
upper off-diagonal elements in $V$, namely,
$|V_{us}|,|V_{ub}|,|V_{cb}|$ as the initial input variables in our analysis,
$i.e.$ the case with Set--A.

If we are given the three input values
of Set--A, then we get the values $|V_{ud}|$ and $|V_{tb}|$:
\begin{eqnarray}
|V_{ud}|^2&=&1-|V_{us}|^2-|V_{ub}|^2 ,
\label{eqvud}   \\
|V_{tb}|^2&=&1-|V_{ub}|^2-|V_{cb}|^2 .
\label{eqvtb}
\end{eqnarray}
To obtain four remaining
elements,$|V_{cd}|,|V_{cs}|,|V_{td}|$ and $|V_{ts}|$, we  write
four constraints for these four elements in Eq.~(\ref{WUC1}) as a
matrix  form:
\begin{equation}
R X = B  ,
\label{con2}
\end{equation}
where
\begin{eqnarray}
R  &=&
\left(
\begin{array}{cccc}
1  &  1  &  0  &  0  \\
0  &  0  &  1  &  1  \\
1  &  0  &  1  &  0  \\
0  &  1  &  0  &  1
\end{array}
\right)  ,
\\
X&=&(|V_{cd}|^2,|V_{cs}|^2,|V_{td}|^2,|V_{ts}|^2)^T ,
\label{varX}
\\
B&=&(1-|V_{cb}|^2,1-|V_{tb}|^2,1-|V_{ud}|^2,1-|V_{us}|^2)^T .
\label{varB}
\end{eqnarray}
In Eqs.~(\ref{varX}),~(\ref{varB}), $X$ and $B$ are column vectors
and $T$ means transpose of the matrix. Because of det$R$=0,
there is not a unique solution if any. In such a situation
there exists a very powerful set of technique,  known as Singular
Value Decomposition (SVD) method. The details of the method are given in the
Appendix A. Remaining mixing elements are expressed as follows:
\begin{eqnarray}
|V_{cd}|^2 &=& -a + u_1 , \nonumber\\
|V_{cs}|^2 &=&  a + u_2 ,\\
|V_{td}|^2 &=&  a + u_3 , \nonumber\\
|V_{ts}|^2 &=& -a + u_4 , \nonumber
\label{comsol}
\end{eqnarray}
where
\begin{eqnarray}
u_1 &=& \frac{1}{4}( 1+2|V_{us}|^2+ |V_{ub}|^2-2|V_{cb}|^2) , \nonumber\\
u_2 &=& \frac{1}{4}( 3-2|V_{us}|^2- |V_{ub}|^2-2|V_{cb}|^2) , \\
u_3 &=& \frac{1}{4}(-1+2|V_{us}|^2+3|V_{ub}|^2+2|V_{cb}|^2) , \nonumber\\
u_4 &=& \frac{1}{4}( 1-2|V_{us}|^2+ |V_{ub}|^2+2|V_{cb}|^2)  , \nonumber
\end{eqnarray}
and a new variable, `$a$', is introduced as a coefficient attached
to the general solution.
If there is no flavor mixing, we can set
$a=1/4$. The value of `$a$' can be determined from Eq.~(9)
if we know any one value of
$|V_{cd}|,|V_{cs}|,|V_{td}|,|V_{ts}|$.
Constraints of non-negative $|V_{ij}|^2$ are applied for the
range of variable $a$:
$$
a_{min}=\max(-u_2,-u_3),~~~
a_{max}=\min(u_1,u_4).
$$
We note that when three input values $|V_{us}|,|V_{ub}|,|V_{cb}|$ are given,
the moduli squared of remaining four mixing elements
$|V_{cd}|,|V_{cs}|,|V_{td}|,|V_{ts}|$ are
just quadratic functions of parameter $a$. As `$a$' increases,
$|V_{cs}|$ and $|V_{td}|$ increase while $|V_{cd}|$ and
$|V_{ts}|$ decrease. $|V_{ud}|$ and $|V_{tb}|$ are fixed
by the three input values and are independent from the parameter $a$.
And the bounds on the parameter `$a$' will determine the regions
of FUC and WUC, which will be explained later.

As a next step,  we further assume that the mixing matrix $V$ satisfies
full unitary conditions.
Then we have six more constraints:
\begin{eqnarray}
\sum_{j=d,s,b} V_{ij} V^*_{kj} &=& 0, \hspace{1cm}
(i,k)=(u,c),(u,t),(c,t) ,  \nonumber  \\
\sum_{j=u,c,t} V_{ji} V^*_{jk} &=& 0, \hspace{1cm}
(i,k)=(d,s),(d,b),(s,b) .
\label{con3}
\end{eqnarray}
These constraints cannot be represented without introduction of
complex numbers analytically. If we know all the absolute values
of $V$, however, we can express necessary and sufficient
conditions for the constraints, Eqs.~(\ref{con3}), in a geometric way.
Eqs.~(\ref{con3}) give six unitarity triangles corresponding to
each six constraints, and all six triangles have equal area that is
directly related to the Jarlskog's rephasing invariant parameter
$J_{CP}$. If we take one of the constraints Eqs.~(\ref{con3}), for
example,
\[
\sum_{j=u,c,t} V_{jd} V^*_{jb} = 0,
\]
a triangle is composed of three sides with lengths
$|V_{ud}||V_{ub}|,|V_{cd}||V_{cb}|,$ and $|V_{td}||V_{tb}|$,
with a necessary condition
\begin{equation}
|V_{cd}||V_{cb}| \le |V_{ud}||V_{ub}| + |V_{td}||V_{tb}| ,
\label{pheno1}
\end{equation}
where the equality holds in CP conserving case.
For more general
argument, let us rewrite Eq.~(\ref{pheno1}) as follows:
\begin{equation}
l_2 \leq l_1 + l_3  ,  \label{COND2}
\end{equation}
where, as an example,  $l_1 = |V_{ud}||V_{ub}|,l_2 = |V_{cd}||V_{cb}|,$
and $l_3 = |V_{td}||V_{tb}|.$ After taking the square on both sides of
Eq.~(\ref{COND2}) we can rearrange the constraint equation as
follows:
\begin{equation}
f(l_1,l_2,l_3)\equiv 2 l_1^2 l_2^2+ 2 l_2^2 l_3^2 + 2 l_1^2 l_3^2
- l_1^4 - l_2^4 - l_3^4 \geq 0  ,  \label{COND3}
\end{equation}
where we denote newly introduced function  $f$ for later use.
Using the Heron's formula, the square of triangular
area can be rewritten as follows:
\begin{equation}
A^2 =s(s-l_1)(s-l_2)(s-l_3)= \frac{1}{16} f(l_1,l_2,l_3) ,
\label{AREA2}
\end{equation}
where $s=(l_1+l_2+l_3)/2$.
So the necessary condition~(\ref{COND3}) for
the complete triangle means non-negative value of $A^2$. The Jarlskog's
invariant parameter is written as follows:
\begin{equation}
J_{CP}=2A=\frac{1}{2}\sqrt{f(l_1,l_2,l_3)}.  \label{JCP}
\end{equation}
If we expand $f$ in terms of parameter $a$:
\begin{eqnarray}
f &=& -(1-|V_{ub}|^2)^2 a^2   \nonumber \\
  & & +2[|V_{ud}|^2|V_{ub}|^2(|V_{tb}|^2-|V_{cb}|^2)
         (|V_{cb}|^2 u_1-|V_{tb}|^2 u_3)
         (|V_{cb}|^2+|V_{tb}|^2)] a   \nonumber \\
  & & +2|V_{ud}|^2|V_{ub}|^2[|V_{cb}|^2 u_1+|V_{tb}|^2 u_3]
      -(|V_{cb}|^2 u_1 - |V_{tb}|^2 u_3)^2 - |V_{ud}|^4|V_{ub}|^4
      ,
\label{FUNC1a}
\end{eqnarray}
where the function $f$ is quadratic of $a$. We can get the boundaries of
the constraint Eq.~(\ref{COND2}), and denote two roots of the
quadratic equation as $a_{-}$ and $a_{+}(>a_{-})$. Two real roots
and the boundary points in the interval depend on only the three
input values. Nonexistence of real solutions of the quadratic
equation means that the three input values do not allow the FUC.
We note that if we force the mixing
matrix $V$ to be fully unitary, then six triangles from the constraints
Eq.~(\ref{con3}) have the same area, which are the sufficient conditions
for the FUC.

We can  relate the coefficient $a$ to the CP
violating parameter in another representation of mixing matrix
with the FUC. Let us consider the standard
parametrization of the CKM matrix
\begin{equation}
V_{CKM} =
\left(
\begin{array}{ccc}
c_{12}c_{13}  &  s_{12}c_{13}  & s_{13} e^{-i\delta_{13}} \\
-s_{12}c_{23}-c_{12}s_{23}s_{13}e^{i\delta_{13}}  &
 c_{12}c_{23}-s_{12}s_{23}s_{13}e^{i\delta_{13}}  &
 s_{23}c_{13}  \\
s_{12}s_{23}-c_{12}c_{23}s_{13}e^{i\delta_{13}}  &
-c_{12}s_{23}-s_{12}c_{23}s_{13}e^{i\delta_{13}}  &
  c_{23}c_{13}
\end{array}
\right) ,
\label{ckm1}
\end{equation}
where $s_{ij}=\sin{\theta_{ij}},c_{ij}=\cos{\theta_{ij}}$. We
find that the coefficient $a$ is directly related to the
parameters in the standard representation:
\begin{equation}
a=-2s_{12}c_{12}s_{23}c_{23}s_{13}\cos{\delta_{13}}
   -\frac{1}{8}\cos{2\theta_{12}}\cos{2\theta_{23}}
  [-3+\cos{2\theta_{13}}]   .
\label{acp}
\end{equation}

Three angles $\alpha,\beta,\gamma$ of the unitarity triangle,
which characterize CP violation,
are defined as follows:
\begin{eqnarray}
\alpha &=& Arg[-(V_{td} V^*_{tb})/(V_{ud} V^*_{ub})] , \\
\beta  &=& Arg[-(V_{cd} V^*_{cb})/(V_{td} V^*_{tb})] , \\
\gamma  &=& Arg[-(V_{ud} V^*_{ub})/(V_{cd} V^*_{cb})] .
\end{eqnarray}
The sum of those three angles, defined as the
intersections of three lines, would be always equal to 180$^0$,
even though the three lines may not be closed to make a triangle,
$i.e.$ in case that CKM matrix is not unitary at all.
We can also define these quantities from the area of the unitary triangle
and its sides:
\begin{eqnarray}
\sin\beta^{\prime} &=&
\frac{2A}{|V_{td}||V_{tb}||V_{cd}||V_{cb}|} , \\
\sin\gamma^{\prime} &=&
\frac{2A}{|V_{ud}||V_{ub}||V_{cd}||V_{cb}|} ,  \\
\alpha^{\prime} &=& \pi-\beta^{\prime}-\gamma^{\prime} ,
\end{eqnarray}
when the FUC is fully satisfied and the area of the triangles
can be defined from (\ref{AREA2}). Any
experimental data that indicates $\alpha\neq\alpha^{\prime}$ or
$\beta\neq\beta^{\prime}$ or $\gamma\neq\gamma^{\prime}$ means
that three generation quark mixing matrix $V$ is not fully
unitary.

\section{Numerical Results and Discussions}

For given input values of $|V_{us}|,|V_{ub}|,|V_{cb}|$, the
parameter `$a$' is divided into two regions depending on whether the
FUC is satisfied or not. We can divide the range of the parameter
into two by setting $l_1=|V_{ud}||V_{ub}|, l_2=|V_{cd}||V_{cb}|,
l_3=|V_{td}||V_{tb}|$, as an example:
\begin{itemize}
\item
Region I : The maximum among $l_1,l_2,l_3$ is larger than
sum of the other two values. In other words it is not possible to
make any triangle with these three segments. This region is
outside of the interval of $(a_{-},a_{+})$.
\item
Region II : The maximum among $l_1,l_2,l_3$ is smaller than sum of
the other two values. In other words it is possible to make a
unitarity triangle. This region is confined to $(a_{-},a_{+})$.
\end{itemize}
In region I, we cannot define $J_{CP}$. On the contrary we can
define $J_{CP}$ in region II and calculate it with $l_1,l_2,l_3$
as shown in Eq.~(\ref{AREA2}). In general the region II is
surrounded by the region I. Two boundary points of region II
correspond to the case of CP conserving case.

For numerical analyses, we refer to the Particle Data
Group (PDG)~\cite{PDB}. Current values of three input
moduli and corresponding sources of measured
matrix elements are summarized in Table I.
\begin{table}
\caption{Input values of the matrix
elements and their sources referred from the PDG. The output
values are allowed intervals (95\% CL) for WUC and FUC.}
\begin{tabular}{lllll}
  &matrix elements & PDG values & Sources & \\  \cline{1-5}
& $|V_{us}|$    &  $0.2196\pm 0.0026$ & $K_{e3}$ decays &   \\
 \cline{2-5}
Input & $|V_{ub}|$    & $(3.6\pm 0.7)\times 10^{-3} $ & $B$ semileptonic
 decays &
 \\\cline{2-5}
 & $|V_{cb}|$    & $(41.2\pm 2.0)\times 10^{-3} $ & $B$ semileptonic
 decays  &
 \\ \hline \hline
 &matrix elements & WUC & FUC & PDG  \\ \cline{1-5}
 & $|V_{cd}|$        &  $0.210 \sim 0.224 $ & $0.214 \sim 0.224 $
 & $0.219 \sim 0.226$  \\
 \cline{2-5}
 & $|V_{cs}|$    & $ 0.9735 \sim 0.9768 $ & $0.9735 \sim 0.9760  $
 & $0.9732 \sim 0.9748$
 \\\cline{2-5}
Output & $|V_{td}|$    & $0.004 \sim 0.045 $ & $ 0.004 \sim 0.014  $
 & $0.004 \sim 0.014$
 \\ \cline{2-5}
  & $|V_{ts}|$    & $ 0.001 \sim 0.045$ & $ 0.035 \sim 0.045 $
 & $0.037 \sim 0.044 $
 \\
\end{tabular}
\end{table}
The input values of $|V_{us}|,|V_{ub}|,|V_{cb}|$ are randomly
generated within 95\% CL with uniform distributions.
Each input determines both two regions for the WUC and FUC.
The WUC is confined to the interval $(a_{min},a_{max})$ of which
calculations are described in the previous section.
The FUC is confined into the interval $(a_{-},a_{+})$ which is
a subset of $(a_{min},a_{max})$.
In the restricted regions we again generated randomly the values
of parameter $a$ for our numeric calculations. Fig. 1(a) shows
scattered points for $|V_{td}|$ and $|V_{ts}|$ values when we apply
the WUC to the choice of parameter $a$. The scattered points
compose a quadrant in the $|V_{td}|$--$|V_{ts}|$ plane. In the figure
we also draw the curved axis for the parameter $a$. The labels on
the curve are valid only when we set the three inputs to
the center values in Table I. If other input values are taken,
the numeric labelling should be slightly changed.
Fig. 1(b) presents scattered points when we apply the FUC
to the choice of parameter $a$. The allowed region for FUC is
much narrower than that for the WUC and is included in the
region for the WUC. The curved axis for the parameter $a$ is
identical to that in Fig. 1(a).
\begin{figure}
    \begin{center}
    \leavevmode
    \epsfysize=8.0cm
    \epsffile[75 160 575 530]{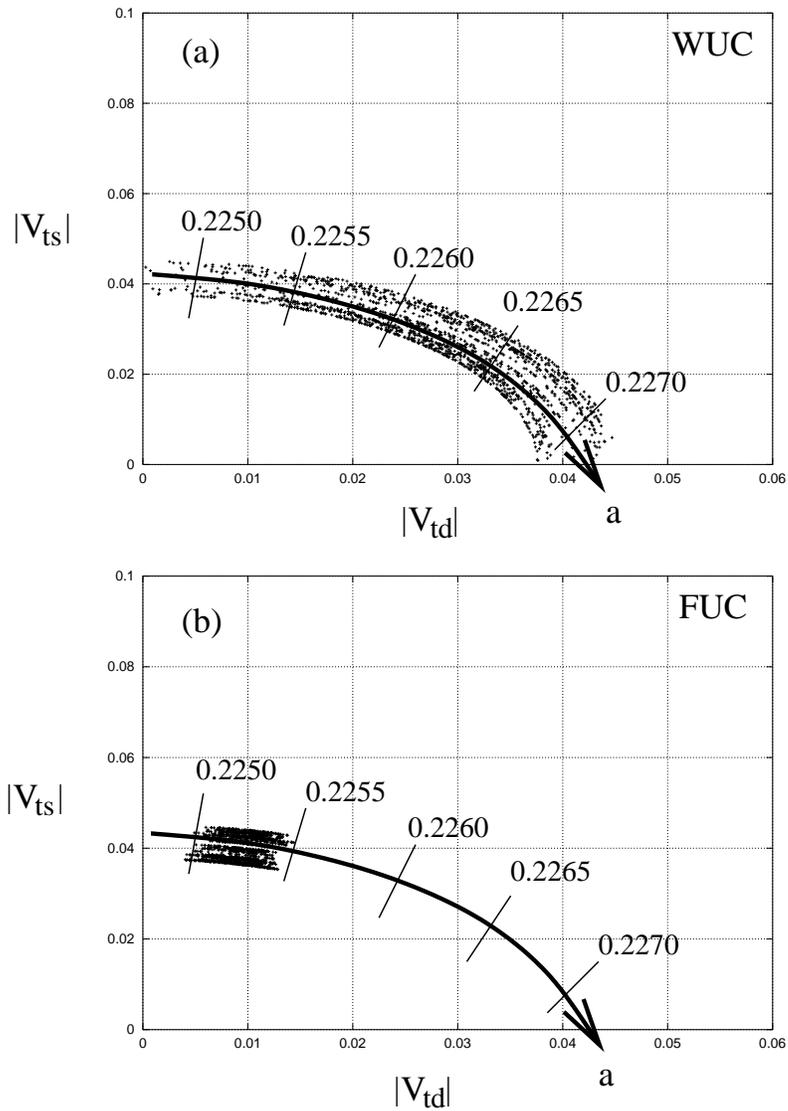}
    \vspace{5.0cm}
    \caption{\label{fig:fig1}
The scattered plots for $|V_{td}|$ and $|V_{ts}|$
values which satisfy the WUC (a) and FUC (b) in the case of
Set--A. The allowed points are calculated from uniformly
generated three input values in 95\% CL. The curved axis for
the parameter `$a$' is drawn with the centered input values
in Table I, not with the randomly generated input values.}
\end{center}
\end{figure}
\begin{figure}
    \begin{center}
    \leavevmode
    \epsfysize=8.0cm
    \epsffile[75 160 575 530]{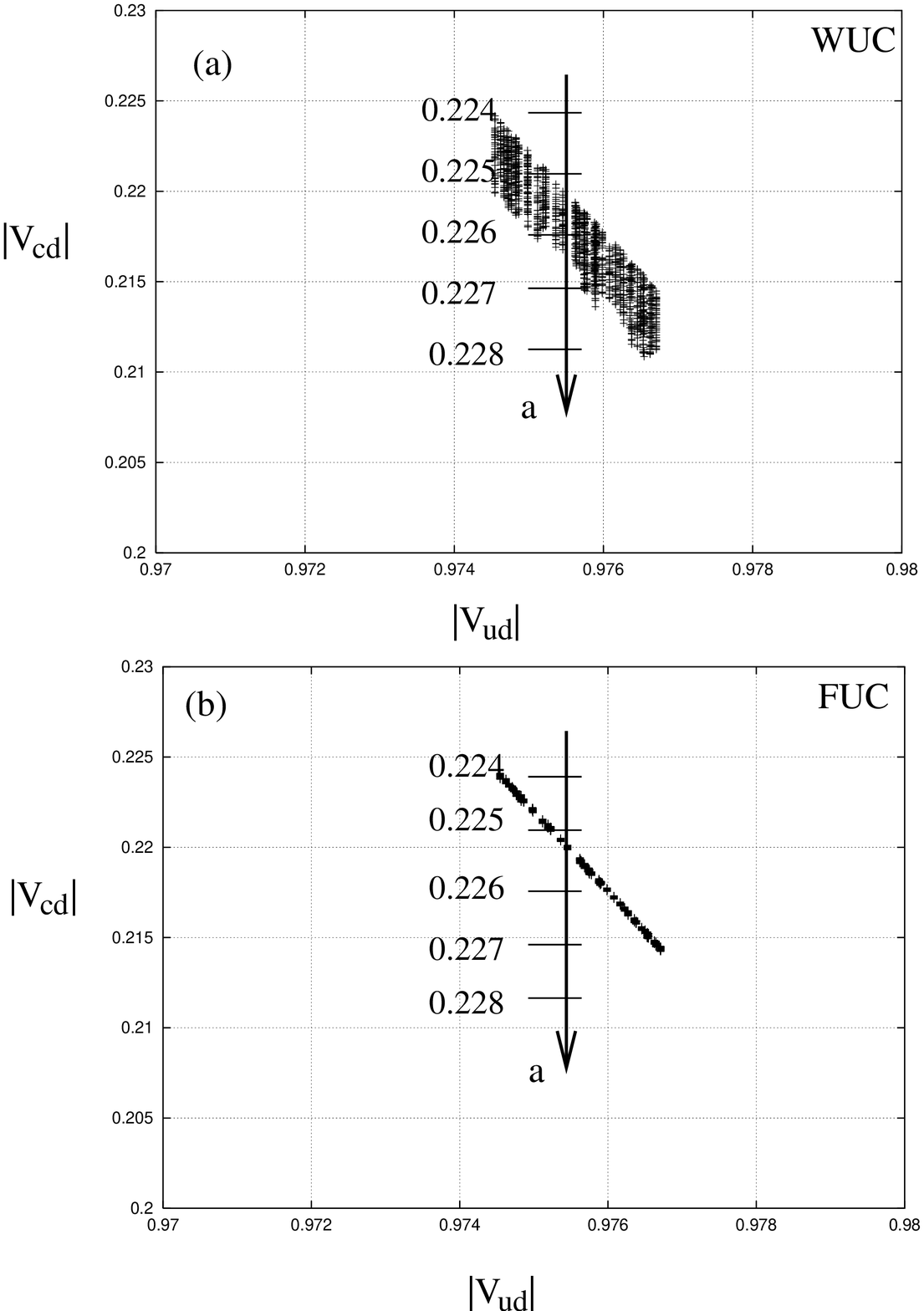}
    \vspace{5.0cm}
    \caption{\label{fig:fig2}
The scattered plots for $|V_{ud}|$ and $|V_{cd}|$
values which satisfy the WUC (a) and FUC (b) in the case of Set--A.
Two panels are corresponding to two panels in Fig.  1.}
\end{center}
\end{figure}
Fig. 2 shows corresponding scattered points for $|V_{ud}|$ and
$|V_{cd}|$ when we take the WUC and FUC.
The moduli $|V_{ud}|$  does not depend on
the parameter $a$ and the $|V_{cd}|$ is directly related to the
value of $a$. The axis for the parameter $a$ is, therefore, a
vertical line along the axis of $|V_{cd}|$.

In Table I, we  show the numerical output values for the moduli,
$|V_{cd}|,~|V_{cs}|,~|V_{td}|,~|V_{ts}|$, within 95\% CL with
the unitary conditions, WUC and FUC. For comparison, we also show
the results of the PDG values.
The PDG values of individual matrix elements were
determined from three-level constraints from weak decays of
the relevant quarks or from deep inelastic neutrino scattering
together with the assumptions of 3 generation FUC.
As can be seen, the allowed regions
for the WUC are much broader than those for the FUC and the
latter are subsets of the former. Our numerical results are
consistent with 90\% CL on the magnitudes of mixing elements
in PDG:
$|V_{td}|\simeq 0.004\sim 0.014$ and
$|V_{ts}|\simeq 0.037\sim 0.044$, particularly. We can see
that the FUC and hierarchical input values of
$|V_{us}|,|V_{ub}|,|V_{cb}|$ imply $|V_{td}|<|V_{ts}|$.
This contrasts with the results of the WUC. In this case,
it is possible that $|V_{td}|$ is equal to or even larger
than $|V_{ts}|$.


If we start with different mixing elements, like
$|V_{ud}|,|V_{us}|,|V_{cd}|$ ($i.e.$ Set--B), then $|V_{td}|$
and $|V_{ub}|$ are first fixed and we introduce new parameter $b$ from
the SVD method, as shown in Appendix B in detail. Remaining four moduli,
$|V_{cs}|,|V_{cb}|,|V_{ts}|$, and $|V_{tb}|$ are dependent on the
parameter $b$ and have correlated values one another.
Precise measurements for one of the four moduli
will fix remaining three moduli.
Similarly, if we start with $|V_{us}|,|V_{cs}|,|V_{cb}|$ ($i.e.$ Set--C), then
$|V_{cd}|$ and $|V_{ts}|$ are fixed and new
parameter $c$ will be introduced by following the SVD method. Remaining
four moduli, $|V_{ud}|,|V_{ub}|,|V_{td}|$ and $|V_{tb}|$ depend on
the parameter $c$ and have correlated values one another.
However, as explained in Section II, it would be much more difficult
numerically to analyze with Set--B or C compared to Set--A
due to approximate unitarity of the upper-left  $2 \times 2$  part
of CKM matrix.

To conclude, we proposed a flexible method in which the unitarity of
quark-mixing matrix can be tested step-by-step. The
singular-value-decomposition (SVD) method is used in analyzing the
mixing matrix over a broader parameter space than the unitary
region as well as in presenting a new parametrization of the CKM matrix.
The question whether the mixing matrix
satisfies the WUC or FUC is a quite difficult and complicate matter
within the standard PDG parametrization or similar unitary parametrization.
In the parametrization by the SVD method the CKM
matrix is represented with three moduli and an additionally
induced flexible parameter $a$\footnote{
In principle, the two methods, 4VKM and SVD, can give
identical results.
However,  in practise the number of input parameters which
should be supplied by experiments is reduced from four to three in SVD.
This reduction in number of {\it a priori} experimental input
values gives conceptually much better way to analyze the CKM mixing
matrix. With the reduced number of input parameters we can check
the consistency between independently measured CKM mixing matrix
elements systematically, and can investigate the inter-relations
among the mixing elements by varying single parameter.}.
Once the value of the induced parameter $a$ is determined, we can
easily distinguish the FUC from the WUC.
For example, with the Set--A input data, if we get
$0.224\le a\le 0.226$, the mixing matrix satisfies the FUC
within 95\% CL. From Fig. 1(b) we can also conclude that the FUC
violated if $|V_{td}|\ge0.02$ or $|V_{td}|\le0.004$.
If $|V_{td}|\ge0.05$, even the WUC is not satisfied.
Fig. 2(b) shows that there is a strong correlation between
$|V_{ud}|$ and $|V_{cd}|$.
This method can also be  applied to the analysis of lepton
flavor mixing, in which only a few moduli are presently measured.

\acknowledgments
We thank A. Dighe for careful reading of the manuscript and his
valuable comments.
The work of C.S.K. was supported
in part by  CHEP-SRC Program,
in part by Grant No. R02-2002-000-00168-0 from BRP of the KOSEF,
in part by BK21 Program and Grant No. 2001-042-D00022 of the KRF.
The work of J.D.K was
supported by the Korea Research Foundation Grant (2001-042-D00022).

\begin{appendix}
\section{The Singular-Value-Decomposition Method}
For detailed description of this method we consider
the specific case with the input parameters,
$|V_{us}|,|V_{ub}|,|V_{cb}|$, which is in Sec. II.
In this case we have to solve Eq.~(\ref{con2}).
According to the method, the matrix $R$ can be decomposed
as a product of three matrices:
\begin{equation}
R=U W V^T  ,
\label{decom}
\end{equation}
where
\begin{eqnarray}
U  &=&
\frac{1}{2}
\left(
\begin{array}{cccc}
-1  &  1  &  1  & -1  \\
-1  & -1  & -1  & -1  \\
-1  &  1  & -1  &  1  \\
-1  & -1  &  1  &  1
\end{array}
\right)  ,
\\
W  &=&
\left(
\begin{array}{cccc}
2  & 0         &  0        & 0  \\
0  & \sqrt{2}  &  0        & 0  \\
0  & 0         & \sqrt{2}  & 0  \\
0  & 0         & 0         & 0
\end{array}
\right)  ,
\\
V  &=&
\frac{1}{2}
\left(
\begin{array}{cccc}
-1  &  \sqrt{2}  &  0         &  1  \\
-1  &  0         &  \sqrt{2}  & -1  \\
-1  &  0         & -\sqrt{2}  & -1  \\
-1  & -\sqrt{2}  &  0         &  1
\end{array}
\right)  .
\end{eqnarray}
In general the matrices $U$ and $V$ are orthogonal in the sense
that their columns are orthonormal,
\begin{eqnarray}
\sum_{i} U_{ik} U_{in}&=&\delta_{kn} ,  \\
\sum_{i} V_{ik} V_{in}&=&\delta_{kn}  .
\end{eqnarray}
We note that this decomposition is not unique.
For further practical calculations of SVD
we refer to Ref.~\cite{NR}.
The solutions of Eq.~(\ref{con2}) are obtained in two types,
special solution and general solution, which can
get in two different ways.
First, the special solution according to SVD is calculated
by defining the {\it inverse} of $R$ as follows:
\begin{eqnarray}
\overline{R} &=& V [diag(1/w_{ii})] U^T  \\
  &=&
  \frac{1}{8}
  \left(
  \begin{array}{cccc}
  3  &  -1  &   3  &  -1  \\
  3  &  -1  &  -1  &   3  \\
  -1 &   3  &   3  &  -1  \\
  -1 &   3  &  -1  &   3
  \end{array}
  \right)  ,
\label{invR}
\end{eqnarray}
where we take $1/w_{ii}=0$ if $w_{ii}=0$.
The matrix $\overline{R}$ is unique and does not
depend on the way how the matrix $R$ is decomposed.
The special solution $X_s$ is
\begin{equation}
X_s = \overline{R} B  .
\label{solxs}
\end{equation}
The inverse matrix $\overline{R}$ does not satisfy the constraints
which must be obeyed in the general sense of the inverse,
namely,
\[
R \overline{R} \neq \overline{R} R \neq I.
\]
However it satisfies
\[
R \overline{R} B = B.
\]
Therefore, we can introduce  general solutions $X_g$
such that
\begin{equation}
R X_g = 0 ,
\label{eqxg}
\end{equation}
and we can add this to the special solution $X_s$.
We can see the general solutions of Eq.~(\ref{eqxg})
by simple guessing as follows:
\begin{equation}
X_g = a (-1,1,1,-1)^T ,
\label{solxg}
\end{equation}
where coefficient $a$ can take any real value. Actually the
coefficient $a$ must be further confined in such a way that the
values of the mixing elements $V_{ij}$ should be within the range
$[0,1]$. We can express the complete solutions as
\begin{equation}
X =   X_g + X_s .
\end{equation}
In algebraic terms, Eq.~(\ref{con2}) defines $R$ as a linear
mapping from the vector space of $X$ to the vector space of $B$.
If $R$ is singular, then there is some subspace of $X$, called the
nullspace, that is mapped to zero,$R X=0$. The number of linearly
independent vectors that can be found in Eq.~(\ref{eqxg}) is the
dimension of the nullspace called the nullity of $R$. In three
generation quark mixing case the nullity is 1.

\section{Case with Set--B input parameters}

If we take three independent input parameters as
$|V_{ud}|,|V_{us}|,$ and $|V_{cd}|$, we can apply the same
procedure to obtain remaining mixing elements. In this case the
matrix Eq.~(\ref{con2}) becomes $RX=B$ with
\begin{eqnarray}
R  &=& \left(
\begin{array}{cccc}
1  &  0  &  1  &  0  \\
0  &  1  &  0  &  1  \\
1  &  1  &  0  &  0  \\
0  &  0  &  1  &  1
\end{array}
\right)  ,
\\
X&=&(|V_{cs}|^2,|V_{cb}|^2,|V_{ts}|^2,|V_{tb}|^2)^T ,
\label{varX2}
\\
B&=&(1-|V_{us}|^2,1-|V_{ub}|^2,1-|V_{cd}|^2,1-|V_{td}|^2)^T .
\label{varB2}
\end{eqnarray}
Two obvious relations from the weak unitary conditions are:
\begin{eqnarray}
|V_{ub}|^2 &=& 1 - |V_{ud}|^2 - |V_{us}|^2 ,  \\
|V_{td}|^2 &=& 1 - |V_{ud}|^2 - |V_{cd}|^2 .
\end{eqnarray}
Following the procedure described in the Appendix A, we can write
total solution of $X$ as follows:
\begin{eqnarray}
|V_{cs}|^2 &=&  b + w_1  ,  \\
|V_{cb}|^2 &=& -b + w_2  ,  \\
|V_{ts}|^2 &=& -b + w_3  ,  \\
|V_{tb}|^2 &=&  b + w_4  , \label{comsol2}
\end{eqnarray}
where
\begin{eqnarray}
w_1 &=& \frac{1}{4}( 3 - |V_{ud}|^2 - 2|V_{us}|^2 - 2|V_{cd}|^2) ,  \\
w_2 &=& \frac{1}{4}( 1 + |V_{ud}|^2 + 2|V_{us}|^2 - 2|V_{cd}|^2) ,  \\
w_3 &=& \frac{1}{4}( 1 + |V_{ud}|^2 - 2|V_{us}|^2 + 2|V_{cd}|^2) ,  \\
w_4 &=& \frac{1}{4}(-1 + 3 |V_{ud}|^2 + 2|V_{us}|^2 + 2|V_{cd}|^2)
,
\end{eqnarray}
and the $b$ is newly introduced parameter. In this case, if there
is no flavor mixing, we can set $b=1/2$. Eq.~(\ref{comsol2}) shows
that any values of $|V_{cs}|,|V_{cb}|,|V_{ts}|,|V_{tb}|$ will
determine the value $b$. Further constraints are applied for the
range of parameter $b$ by $|V_{ij}|^2 \ge 0$:
$b_{min}=max(-w_1,-w_4), b_{max}=min(w_2,w_3)$. For the FUC we
expand $f$ in terms of parameter $b$:
\begin{eqnarray}
f &=& -(1-|V_{ud}|^2)^2 b^2   \nonumber \\
  & & +2[|V_{ud}|^2|V_{ub}|^2(|V_{td}|^2-|V_{cd}|^2)
         (|V_{cd}|^2 w_2-|V_{td}|^2 w_4)
         (|V_{cd}|^2+|V_{td}|^2)] b  \nonumber \\
  & & +2|V_{ud}|^2|V_{ub}|^2[|V_{cd}|^2 w_2+|V_{td}|^2 w_4]
      -(|V_{cd}|^2 w_2 - |V_{td}|^2 w_4)^2 - |V_{ud}|^4|V_{ub}|^4 .
\label{FUNC1b}
\end{eqnarray}
Like the previous case we can get the boundaries of the constraint
Eq.~(\ref{COND2}), and denote two roots of the quadratic equation
as $b_{-}$ and $b_{+}(>b_{-})$.

\end{appendix}


\end{document}